\documentclass[11pt]{article}

\usepackage{amsmath,amssymb,amscd,amsfonts}
\allowdisplaybreaks[4]

\begin{document}

\begin{flushright} 
 
WIS/14/10-AUG-DPPA\\

\end{flushright} 

\vspace{0.1cm}

\begin{center}
  {\LARGE
 A proposal of a fine tuning free formulation of \\
4d ${\cal N}=4$ super Yang-Mills
  }
\end{center}
\vspace{0.1cm}
\vspace{0.1cm}
 \begin{center}

          Masanori H{\sc anada}\footnote
          {
 E-mail address : masanori.hanada@weizmann.ac.il} 
\\
\vspace{5mm}
Department of Particle Physics and Astrophysics\\
Weizmann Institute of Science\\
Rehovot 76100, Israel   

 \end{center}
\vspace{1.5cm}

\begin{center}
  {\bf Abstract}
\end{center}

Recently, a nonperturbative formulation of 4d ${\cal N}=4$ 
super Yang-Mills theory which does not 
require fine tuning at least to all order in perturbation theory 
has been proposed by combining two-dimensional lattice and matrix model techniques. 
In this paper we provide an analogous model by utilizing deconstruction approach 
of Kaplan et al. 
Two-dimensional lattice with a plane wave deformation is deconstructed 
from a matrix model and two additional dimensions emerge through the Myers effect. 
In other words we construct a D1-brane theory from which a D3-brane theory comes out.   
The action is much simpler than the previous formulation 
and hence numerical study, 
which enables us to test the $AdS_5/CFT_4$ duality 
at fully nonperturbative level, becomes much easier.

\newpage


\section{Introduction}
Supersymmetric Yang-Mills (SYM) theories play prominent roles in theoretical particle physics. 
Among them, maximally supersymmetric theories are of crucial importance 
for superstring/M theory \cite{BFSS,IKKT,MatrixString,Maldacena:1997re}. 
Given that most interesting questions can be answered only through nonperturbative study, 
it is important to construct theoretical frameworks for that. 
However, it is not a straightforward task because of the notorious 
difficulties of lattice supersymmetry (SUSY). 
So far, lattice formulations 
which are free from fine tunings are established only for one- and two-dimensional 
theories, three-dimensional maximally supersymmetric theory 
and 4d ${\cal N}=1$ pure SYM.\footnote{
For recent numerical studies of 4d ${\cal N}=1$ theory, see \cite{Giedt:2008xm}. 
For review of orbifold method which is utilized in this paper, see \cite{Catterall:2009it}. 
Note however that a part of numerical data for super Yang-Mills theories 
shown in this review is problematic; see \cite{Hanada:2010qg}. 
}
It motivated people to study {\it non-lattice} approaches to SYM. 

For 1d theory (matrix quantum mechanics), lattice is not necessary at all 
thanks to the absence of UV divergence and simple 
momentum cutoff prescription works \cite{Hanada-Nishimura-Takeuchi}.  
By using it, remarkable 
quantitative agreement with the gauge/gravity duality conjecture 
\cite{Maldacena:1997re} has been obtained \cite{AHNT}
(for lattice study with qualitatively consistent result, see \cite{Catterall-Wiseman}). 
By combining the momentum cutoff or lattice techniques   
with a plane wave deformation \cite{BMN} and the Myers effect \cite{Myers:1999ps}, 
3d theory can be obtained 
as an expansion of 1d matrix model around fuzzy sphere \cite{Maldacena:2002rb}. 
Also, in the planar limit, 4d theory 
can be obtained using a novel large-$N$ reduction technique \cite{Ishii:2008ib,Hanada:2009hd}
inspired by the Eguchi-Kawai equivalence \cite{Eguchi:1982nm}. 

In order to construct 4d ${\cal N}=4$ SYM at a finite rank of a gauge group, 
one can combine fuzzy sphere technique \cite{Maldacena:2002rb} with 2d SYM; 
that is, by constructing 2d SYM with the plane wave deformation using standard lattice SUSY techniques 
and then taking fuzzy sphere background, 4d SYM is naturally realized. Such a model 
is constructed in \cite{Hanada:2010kt} by modifying Sugino's 2d lattice model 
\cite{Sugino:2003yb,Sugino:2004qd,Sugino:2004uv}, 
and the absence of fine tuning problem to all order in perturbation theory  
has been shown. Whether fine tunings are absent at nonperturbative level should be checked 
by numerical simulation. However note that in other models the absence is not shown 
even at perturbative level\footnote{
In a class of lattice models, fine tuning is not needed at one-loop level 
\cite{Catterall_private}. It would be nice to study whether it is the case 
at higher order. 
}. 

Although this model possesses beautiful features, however, the action is rather complicated 
and it is not easy to put it on computer. Therefore, in this paper we construct similar, 
but much simpler, model by utilizing the deconstruction method (or ``deconstruction/orbifolding 
approach") of Kaplan et al. 
\cite{Kaplan:2002wv,Cohen:2003xe,KaplanUnsal16SUSY}\footnote{
For other constructions, see \cite{Catterall:2004np,D'Adda:2005zk,Damgaard:2007be}. 
Relationship between various models are discussed in \cite{Unsal:2006qp}. 
} . 
The action is as simple as the original non-deformed model \cite{KaplanUnsal16SUSY}  
and can easily be put on computer.

\section{Basic idea}
First let us remind a matrix model construction of 3d maximally 
supersymmetric Yang-Mills theory \cite{Maldacena:2002rb};  
in short, {\it D2-brane theory (3d SYM) emerges from D0-brane theory (1d SYM)}.  
The starting point is the $U(N)$ plane wave matrix model of Berenstein-Maldacena-Nastase 
\cite{BMN}. Fuzzy sphere solution to this model, 
which is interpreted as compact fuzzy D2-branes, 
preserves 16 SUSY.  
Around $k$-coincident fuzzy sphere 3d $U(k)$ theory (D2-brane theory) on noncommutative space is realized.   
At any fixed $N$, this is nothing but 1d theory and can easily be regularized.  
By taking the continuum limit as 1d theory, full SUSY is restored automatically, 
without requiring any fine tuning. {\it By taking continuum limit first for 1d (time) direction 
and then along spherical directions, 3d SYM is realized without parameter fine tuning} \footnote{
Similar anisotropic continuum limit is considered also in the framework 
of the deconstruction in order to reduce the number of fine tunings \cite{Kaplan:2002wv}. 
}. 
Note that for the maximally supersymmetric theory commutative limit of 
noncommutative space is expected to be smooth \cite{Hashimoto:1999ut,Matusis:2000jf}.

To obtain 4d theory, one can start with 2d $U(N)$ SYM with the plane wave deformation 
\cite{Das:2003yq}; {\it D3-brane theory (4d SYM) is obtained from D1-brane theory (2d SYM)}\footnote{
Another possible option is to consider four-dimensional noncommutative space out of 
zero-dimensional theory. For example in \cite{Unsal:2005us} and \cite{Ydri:2007ua} 
realization by using fuzzy four-torus and fuzzy $S^2\times S^2$ have been discussed. 
However these geometries have flat or tachyonic directions \cite{Bal:2004ai} 
and it is not clear 
whether the geometry can be stabilized by adding soft deformations.  
Other geometries like fuzzy ${\mathbb C}P^2$  
might be useful \cite{Kaneko:2005kp}. 
Note also that the boundary condition of the fermions cannot be changed in 
these constructions and hence thermal properties cannot be studied. 
}.   
Four-dimensional $U(k)$ theory on two commutative and 
two noncommutative dimensions naturally arises around $k$-coincident fuzzy sphere background. 
By taking an appropriate large-$N$ limit, 4d ${\cal N}=4$ SYM on {\it commutative} 
${\mathbb R}^4$ is obtained \cite{Hanada:2010kt}. 
(This point is further explained in \S~\ref{sec:4d}.)  

In \cite{Hanada:2010kt}, such 2d lattice is constructed by generalizing Sugino model \cite{Sugino:2003yb,Sugino:2004qd,Sugino:2004uv}. 
However the action is rather complicated and 
cannot easily be put on computer. 
Actually already before turning on the plane wave deformation
Sugino's action is much more complicated than ``deconstruction" model 
of Kaplan et al. \cite{Kaplan:2002wv,Cohen:2003xe,KaplanUnsal16SUSY}; 
so we repeat the program pursued in \cite{Hanada:2010kt}, this time using 
the deconstruction technique, and construct a simple action which is convenient 
for numerical simulation. 

We start with a zero-dimensional maximally supersymmetric matrix model (IIB matrix model) \cite{IKKT}, 
which was used as a starting point in \cite{KaplanUnsal16SUSY}.  
To IIB matrix model we add ``plane-wave deformation", 
by utilizing the results obtained in \cite{Hanada:2010kt}. 
From that we construct 2d theory following the procedure of Kaplan et al. 
It turns out that the procedure in \cite{KaplanUnsal16SUSY} applies perfectly in parallel; 
Although we have a small number of deformation terms, 
essentially the same orbifolding condition can be used.  

In summary, there are three basic steps: 
\begin{itemize}
\item
Add supersymmetric ``plane wave" deformation to IIB matrix model. (\S~\ref{sec:mother})

\item
From the matrix model, generate 2d lattice through deconstruction of Kaplan et al. 
Then take continuum limit. (\S~\ref{sec:deconstruction})

\item
From 2d SYM, generate 4d SYM through the Myers effect. (\S~\ref{sec:4d})

\end{itemize}

\section{Brief introduction to the deconstruction}
In this section we provide a short review of the deconstruction method 
\cite{ArkaniHamed:2001ca}. 
Although we explain only two-dimensional lattice, generalizations to 
other dimensions are straightforward. 

Let us start with bosonic 4-matrix model, 
which is obtained from 4d pure Yang-Mills theory through the dimensional reduction  
(``{\it mother theory}") 
\begin{eqnarray} 
S
=
-\frac{1}{4g_{0d}^2}Tr
[X_I,X_J]^2, 
\end{eqnarray}
where $X_I (I=1,\cdots,4)$ are $N\times N$ hermitian matrices. 
We take $N=L^2M$, where $L$ and $M$ are integers as well. 
$L$ translates into the size of lattice, while $M$ specifies 
the gauge group $U(M)$. 

By using complex fields $x\equiv X_1+iX_2$ and $y\equiv X_3+iX_4$, 
the action can be written as 
\begin{eqnarray}
S
=
\frac{1}{4g_{0d}^2}Tr
\left\{
\frac{1}{2}
|[x,\bar{x}]+[y,\bar{y}]|^2
+
2|[x,y]|^2
\right\}. 
\end{eqnarray}
Here $\bar{x}=X_1-iX_2$ and $\bar{y}=X_3-iX_4$. 

In the deconstruction method, two-dimensional lattice (``{\it daughter theory}")
is obtained from the matrix model (mother) through the orbifolding. 
For fields $\Pi (=x,y)$, we introduce ``orbifolding condition" 
\begin{eqnarray}
C_i\Pi C_i^{-1}=\omega^{r_{\Pi}^i}\Pi, 
\qquad
\omega=e^{2\pi i/L}
\end{eqnarray}
where $C_1$ and $C_2$ are given by 
\begin{eqnarray}
C_1=\Omega\otimes\textbf{1}_L\otimes\textbf{1}_M, 
\qquad
C_2=\textbf{1}_L\otimes\Omega\otimes\textbf{1}_M,
\end{eqnarray}
and 
\begin{eqnarray}
\Omega
=
diag(\omega^{-1},\omega^{-2},\cdots,\omega^{-L})
\end{eqnarray}
is the clock matrix. 
Here we take 
\begin{eqnarray}
\vec{r}_x=(1,0), 
\qquad
\vec{r}_y=(0,1).   
\end{eqnarray} 

Then, the only non-vanishing components are $x_{n_1,n_2,k;n_1+1,n_2,l}$ and 
$y_{n_1,n_2,k;n_1,n_2+1,l}$ ($n_1,n_2=1,\cdots,L;k,l=1,\cdots,M$). 
We interpret $(n_1,n_2)$ to be a label of site on $L\times L$ 
periodic lattice.   
Also we regard $x$ and $y$ as link variables connecting $(n_1,n_2)$ and 
$(n_1+1,n_2)$, $(n_1,n_2+1)$, respectively. 
By denoting  $x_{\vec{n},kl}\equiv x_{n_1,n_2,k;n_1+1,n_2,l}$ and 
$y_{\vec{n},kl}\equiv y_{n_1,n_2,k;n_1,n_2+1,l}$, one obtains
\begin{eqnarray}
S^{lat}
&=&
\frac{1}{4g_{0d}^2}\sum_{\vec{n}}Tr\Bigl\{
\frac{1}{2}|
x_{\vec{n}}\bar{x}_{\vec{n}}
-\bar{x}_{\vec{n}-\hat{x}}x_{\vec{n}-\hat{x}}
+
y_{\vec{n}}\bar{y}_{\vec{n}}
-\bar{y}_{\vec{n}-\hat{y}}y_{\vec{n}-\hat{y}}
|^2
+
2|
x_{\vec{n}}y_{\vec{n}+\hat{x}}
-
y_{\vec{n}}x_{\vec{n}+\hat{y}}
|^2
\Bigl\}. 
\end{eqnarray}
Here $Tr$ indicates the trace for $M\times M$ matrix. 
By expanding it around 
\begin{eqnarray}
x_{\vec{n}}
=
\frac{1}{a}+s_{1,\vec{n}}+iA_{1,\vec{n}}, 
\qquad
y
=
\frac{1}{a}+s_{2,\vec{n}}+iA_{2,\vec{n}},  
\end{eqnarray}
and by taking $g_{2d}^2 \equiv a^2g_{0d}^2$, 
tree-level continuum limit becomes 
\begin{eqnarray}
S^{cont,tree}
=
\frac{1}{2g_{2d}^2}\int d^2x
\left(
F_{12}^2
+
(D_\mu s_I)^2
-
[s_1,s_2]^2
\right). 
\end{eqnarray}
Similar expansion can be performed around 
more generic background $x_{\vec{n}}\bar{x}_{\vec{n}}
=y_{\vec{n}}\bar{y}_{\vec{n}}=\frac{1}{a^2}\cdot\textbf{1}$. 

When $x_{\vec{n}}\bar{x}_{\vec{n}}$ and $y_{\vec{n}}\bar{y}_{\vec{n}}$ 
deviate from $\frac{1}{a^2}\cdot\textbf{1}$, 
the continuum limit does not agree with Yang-Mills theory; in terms of $U_i$, 
it corresponds to the situation that the background value of plaquette deviates 
from 1. In order to avoid such pathological situation, one adds 
\begin{eqnarray}
\frac{\nu^2 a^2}{8g_{0d}^2}\sum_{\vec{n}}
Tr\left(
\left|x_{\vec{n}}\bar{x}_{\vec{n}}-\frac{1}{a^2}\right|^2
+
\left|y_{\vec{n}}\bar{y}_{\vec{n}}-\frac{1}{a^2}\right|^2
\right)  
\end{eqnarray} 
to the action. In the continuum, this amounts to adding a scalar mass term 
$(\nu^2/2g_{2d}^2)\int d^2x Tr(s_1^2+s_2^2)$.  

The advantage of the deconstruction method when we consider supersymmetric theory is 
\cite{Kaplan:2002wv} 
{\it if the SUSY transformation generated by a supercharge $Q$ which relates fields 
of the same charge $\vec{r}$, the daughter theory automatically has an exact 
supersymmetry generated by (projected version of) $Q$.} 
This fact is obvious but rather powerful, because supersymmetric matrix model can easily be obtained 
just by dimensional reduction, and an appropriate choice of $\vec{r}$ is 
naturally obtained from R-symmetry charge.  

\section{0d ``mother" theory}\label{sec:mother}

Let us start with the IIB matrix model \cite{IKKT}, which is the zero-dimensional reduction 
of 4d ${\cal N}=4$ SYM. 
It is written as 
\begin{eqnarray}
S_{0}
&=&
\frac{1}{g_{0d}^2}
Tr
\biggl\{ 
-
\frac{1}{4}\left[X^I,X^J\right]^2
+
\frac{i}{2}\Psi^T\gamma_I\left[X^I,\Psi\right]
\biggl\},
\end{eqnarray}
where 
$I=1,\cdots,10$, and 
$\gamma_I$ ($I=1,\cdots,10$) are $16\times 16$ sub-sectors of 10d gamma matrices.   

In order to introduce plane wave deformation, it is useful to switch to 
another (at first sight rather complicated) notation\footnote{
This is obtained from BTFT formulation of 4d ${\cal N}=4$ SYM in \cite{Sugino:2003yb} by 
dimensional reduction. Here, we redefine $H_A+\frac12\epsilon_{ABC}[B_B, B_C]$, $\phi$, $\bar{\phi}$  
in (4.13) in \cite{Sugino:2003yb} as $H_A$, $\phi_+$, $\phi_-$, respectively.   
} by using Hermitian scalars 
$X_i\ (i=1,2,3,4), B_A\ (A=1,2,3)$ and $C$, complex scalars 
$\phi_\pm$, 
bosonic auxiliary fields $H_A$, 
$\tilde{H}_i$,   
and fermionic variables $\psi_{\pm i}, \chi_{\pm A}$ and $\eta_{\pm}$.  
There are appropriate supercharges $Q^{(0)}_{\pm}$ by which
$S_0$ can be written in exact form as   
\begin{eqnarray} 
S_0
=
Q^{(0)}_+Q^{(0)}_-
{\cal F}^{(0)}, 
\label{Q^{(0)}-closed form}
\end{eqnarray}
where 
\begin{eqnarray}
{\cal F}^{(0)}
&=&
\frac{1}{2g^2_{0d}}
Tr\Bigl\{
-iB_A\Phi_A
-
\frac{1}{3}\epsilon_{ABC}B_A[B_B,B_C]
\nonumber\\
& &
-
\psi_{+i}\psi_{-i}
-
\chi_{+A}\chi_{-A}
-
\frac{1}{4}\eta_+\eta_-
\Bigl\},  
\label{F}
\label{4d N=4 continuum action}
\end{eqnarray}
and $\Phi_1=2(-i[X_1,X_3]-i[X_2,X_4])$, 
$\Phi_2=2(-i[X_1,X_4]+i[X_2,X_3])$,  
$\Phi_3=2(-i[X_1,X_2]+i[X_3,X_4])$. 
Supercharges $Q_{\pm}^{(0)}$ transform fields as  
\begin{eqnarray}
& &
Q^{(0)}_{\pm}X_i=\psi_{\pm i},
\quad
Q^{(0)}_\pm\psi_{\pm i}
=
\mp \left[X_i,\phi_{\pm}\right], 
\nonumber\\
& &
Q^{(0)}_\mp\psi_{\pm i}
=
-\frac{1}{2}[X_i, C]\mp\tilde{H}_i, 
\nonumber\\
& &
Q^{(0)}_{\pm}\tilde{H}_i
=
\left[\phi_{\pm},\psi_{\mp i}\right]
\mp\frac{1}{2}\left[C,\psi_{\pm i}\right]
\pm\frac{1}{2}\left[X_i,\eta_{\pm}\right], 
\nonumber\\
& &
Q^{(0)}_{\pm}B_A
=
\chi_{\pm A}, 
\quad
Q^{(0)}_{\pm}\chi_{\pm A}
=
\pm[\phi_\pm,B_A], 
\nonumber\\
& &
Q^{(0)}_\mp\chi_{\pm A}
=
-\frac{1}{2}[B_A,C]
\mp H_A, 
\nonumber\\
& &
Q^{(0)}_\pm H_A
=
[\phi_\pm,\chi_{\mp A}]
\pm\frac{1}{2}\left[B_A,\eta_\pm\right]
\mp\frac{1}{2}\left[C,\chi_{\pm A}\right], 
\nonumber\\ 
& &
Q^{(0)}_\pm C
=
\eta_\pm, 
\quad
Q^{(0)}_\pm\eta_\pm
=
\pm\left[\phi_\pm,C\right], 
\nonumber\\
& &
Q^{(0)}_\mp\eta_\pm
=
\mp\left[\phi_+,\phi_-\right], 
\nonumber\\
& &
Q^{(0)}_\pm\phi_\pm=0, 
\quad
Q^{(0)}_{\mp}\phi_\pm=\mp\eta_\pm.  
\label{SUSY algebra_4dN=4_continuum}
\end{eqnarray}
One can see the nilpotency 
$\left(Q_+^{(0)}\right)^2 = \left(Q_-^{(0)}\right)^2 
= \{Q_+^{(0)},Q_-^{(0)}\}=0$  
up to gauge transformations. 

We introduce a mass parameter $\mu$ to deform these charges 
as \cite{Hanada:2010kt}\footnote{
This is obtained by dimensionally reducing the algebra in two dimensions \cite{Hanada:2010kt}. 
Note that, in contrast to the plane wave matrix model \cite{BMN}, supersymmetry transformation 
parameter does not depend on the coordinate and hence the dimensional reduction works.  
}
\begin{eqnarray}
Q_{\pm}=Q_\pm^{(0)}+\Delta Q_{\pm}, 
\end{eqnarray}
where non-vanishing $\Delta Q_\pm$ transformations are  
\begin{eqnarray}
& & 
\Delta Q_{\pm}\tilde{H}_i
= 
\frac{\mu}{3}\psi_{\pm i}, 
\qquad
\Delta Q_\pm H_A
= 
\frac{\mu}{3}\chi_{\pm A}, 
\nonumber\\
& & 
\Delta Q_\pm\eta_\pm
= 
\frac{2\mu}{3}\phi_\pm, 
\qquad
\Delta Q_\mp\eta_\pm
= 
\pm\frac{\mu}{3}C. 
\label{SUSY algebra_deformation}
\end{eqnarray}
Then $Q_\pm$ satisfy the anti-commutation relations,  
\begin{eqnarray}
& &
Q_+^2
=
\frac{\mu}{3}J_{++}, 
\quad
Q_-^2
=
-\frac{\mu}{3}J_{--}, 
\nonumber\\
& &
\{Q_+,Q_-\}
=
-\frac{\mu}{3}J_0,  
\label{Q-anticommutator}
\end{eqnarray}
up to gauge transformations, where $J_0$, $J_{++}$ and $J_{--}$ are 
generators of $SU(2)_R$ symmetry~\cite{Sugino:2003yb}. 
The eigenvalues of $J_0$ are $\pm 1$ for the fermions with index $\pm$, 
$\pm 2$ for $\phi_\pm$, and zero for the other bosonic fields. 
Note that $\phi_\pm$ and $C$ form an $SU(2)_R$ triplet and each pair of 
$(\psi_{+i},\psi_{-i})$, 
$(\chi_{+A},\chi_{-A})$, 
$(\eta_+,-\eta_-)$ and $(Q_+,Q_-)$ forms a doublet. 
In particular, $[J_{\pm\pm}, Q_\pm]=0$, $[J_{\pm\pm}, Q_\mp]=Q_\pm$.  

Using the modified supercharges, we can define $Q_\pm$-closed action as 
\begin{eqnarray}
S=
\left(
Q_+Q_- - \frac{\mu}{3}
\right)
{\cal F}, 
\label{Q-closed form}
\end{eqnarray}
where 
\begin{eqnarray}
{\cal F}
&=&
{\cal F}^{(0)}
+
\Delta {\cal F}, 
\nonumber\\
\Delta {\cal F}
&=&
\frac{1}{2g_{0d}^2}\ 
Tr\left(
\frac{1}{2}\sum_{A=1}^3 a_AB_A^2
+
\frac{1}{2}\sum_{i=1}^4 c_i X_i^2 
\right). 
\label{mass_deformation}
\end{eqnarray}
That the action \eqref{Q-closed form} is $Q_\pm$-closed can easily be seen 
 by using \eqref{Q-anticommutator} and the $SU(2)_R$ invariance of ${\cal F}$. 
Here we take $a_A=-\frac{2\mu}{3}$ and $c_i=0$    
for convenience. 
After integrating out auxiliary fields, the action reads 
\begin{eqnarray}
S=S_0+\Delta S, 
\end{eqnarray}
where
\begin{eqnarray}
& &
\Delta S
=
\frac{1}{2g_{0d}^2}Tr
\Bigl\{ 
-\frac{\mu}{2}C[\phi_+,\phi_-]
+
\frac{\mu^2}{9}
\left(
\frac{C^2}{4}+\phi_+\phi_-
\right)
+
\frac{2\mu}{3}\psi_{+i}\psi_{-i} 
-
\frac{\mu}{6}\eta_+\eta_-
-
\frac{4\mu}{3}B_1[B_2,B_3] 
\Bigl\}. 
\nonumber\\
\end{eqnarray}
{}From this expression one can see some similarity 
to the plane wave matrix model 
\cite{BMN}.  Note that the term 
$Tr B_1[B_2,B_3]$ is purely imaginary.

\section{Deconstructing 2d theory}\label{sec:deconstruction}

\subsection{Deconstructing 2d plane-wave SYM}
In this section we construct a two-dimensional lattice through the deconstruction. 

As already mentioned, the advantage of the deconstruction method is 
{\it if the orbifolding commutes with SUSY transformation generated by $Q_\pm$, 
it is promoted to the supersymmetry of the lattice.}  
To find such an orbifolding condition, we introduce complex fields
\begin{eqnarray}
& &
x=X_1+iX_2, 
\qquad
y=X_3+iX_4, 
\nonumber\\
& &
\xi_{\pm x}=\psi_{\pm 1}+i\psi_{\pm 2}, 
\qquad
\xi_{\pm y}=\psi_{\pm 3}+i\psi_{\pm 4},
\nonumber\\
& &
\tilde{h}_{x}=\tilde{H}_{1}+i\tilde{H}_{2}, 
\qquad
\tilde{h}_{y}=\tilde{H}_{3}+i\tilde{H}_{4} 
\end{eqnarray}
and
\begin{eqnarray}
b=B_1+iB_2, 
\qquad
\rho_{\pm}
=
\chi_{\pm 1}+ i\chi_{\pm 2}, 
\qquad
h=H_1+iH_2.  
\end{eqnarray}
Then we assign
\begin{eqnarray}
& &
\vec{r}_x=(1,0)
\qquad
(x,\xi_{\pm x}\ {\rm and}\ \tilde{h}_x), 
\\
& &
\vec{r}_y=(0,1)
\qquad
(y,\xi_{\pm y}\ {\rm and}\ \tilde{h}_y), 
\\
& &
\vec{r}_b=(-1,1)
\qquad
(b,\rho_{\pm}\ {\rm and}\ h).  
\end{eqnarray}
For other fields, $\vec{r}$ is taken to be zero. 
This charge assignment is similar to the one used in 2d ${\cal N}=(8,8)$ Kaplan-\"{U}nsal model 
\cite{KaplanUnsal16SUSY}.  
Compatibility with $Q_\pm$ can easily be seen; 
fields with the same charge mixes linearly, up to multiplication of neutral fields. 
For example, $x, \xi_{\pm x}$ and $\tilde{h}_x$ transform as 
\begin{eqnarray}
& &
Q^{(0)}_{\pm}x=\xi_{\pm x},
\quad
Q^{(0)}_\pm\xi_{\pm x}
=
\mp \left[x,\phi_{\pm}\right], 
\nonumber\\
& &
Q^{(0)}_\mp\xi_{\pm x}
=
-\frac{1}{2}[x, C]\mp\tilde{h}_x, 
\nonumber\\
& &
Q^{(0)}_{\pm}\tilde{h}_x
=
\left[\phi_{\pm},\xi_{\mp x}\right]
\mp\frac{1}{2}\left[C,\xi_{\pm x}\right]
\pm\frac{1}{2}\left[x,\eta_{\pm}\right]. 
\end{eqnarray}
Neutrality of the action follows from the neutrality of ${\cal F}$. It can be seen by rewriting 
${\cal F}$ by complex notation; for example, ${\cal F}^{(0)}$ becomes 
\begin{eqnarray}
{\cal F}^{(0)}
&=&
\frac{1}{2g_{0d}^2}Tr\Bigl\{
-
\bar{b}[\bar{x},y]
-
b[x,\bar{y}]
+
iB_3\left(-[x,\bar{x}]+[y,\bar{y}]\right)
-
iB_3[b,\bar{b}]
\nonumber\\
& &
-
\frac{\bar{\xi}_{+x}\xi_{-x}
+\xi_{+x}\bar{\xi}_{-x}
}{2}
-
\frac{\bar{\xi}_{+y}\xi_{-y}
+\xi_{+y}\bar{\xi}_{-y}
}{2}
-
\frac{\bar{\rho}_{+}\rho_{-}
+\rho_{+}\bar{\rho}_{-}
}{2}
-
\chi_{+3}\chi_{-3}
-
\frac{1}{4}\eta_+\eta_-
\Bigl\},   
\end{eqnarray}
where 
$\bar{\xi}_{\pm x}\equiv\psi_{\pm 1}-i\psi_{\pm 2}$, 
$\bar{\xi}_{\pm y}\equiv\psi_{\pm 3}-i\psi_{\pm 4}$ and 
$\bar{\rho}_{\pm}\equiv\chi_{\pm 1}-i\chi_{\pm 2}$. 
Note that fields with bars have charges of the opposite sign.  

By the deconstruction, fields with charge $(1,0)$ and $(0,1)$ become 
link variables on $x$- and $y$-directions, respectively,   
those with $(-1,1)$ become variables on $(-\hat{x},\hat{y})$ ``diagonal" link, 
and those with $(0,0)$ become site variables.  

In the following we will show only bosonic part. (Fermionic part is shown in appendix.) 
Mother theory is
\begin{eqnarray}
S_{0}^{bos}
&=&
\frac{1}{2g_{0d}^2}Tr\Bigl\{
\frac{1}{4}|[x,\bar{x}]+[y,\bar{y}]|^2
+
|[x,y]|^2
+
\frac{1}{4}|[\phi_+,\phi_-]|^2
+
\frac{1}{4}|[\phi_+,C]|^2
+
\frac{1}{4}|[b,\bar{b}]|^2
+
|[b,B_3]|^2
\nonumber\\
& &
\qquad
+
\frac{1}{2}|[x,\phi_+]|^2
+
\frac{1}{2}|[\bar{x},\phi_+]|^2
+
\frac{1}{4}|[x,C]|^2
+
\frac{1}{2}|[y,\phi_+]|^2
+
\frac{1}{2}|[\bar{y},\phi_+]|^2
+
\frac{1}{4}|[y,C]|^2
\nonumber\\
& &
\qquad
+
\frac{1}{2}|[x,b]|^2
+
\frac{1}{2}|[\bar{x},b]|^2
+
|[x,B_3]|^2
+
\frac{1}{2}|[y,b]|^2
+
\frac{1}{2}|[\bar{y},b]|^2
+
|[y,B_3]|^2
\nonumber\\
& &
\qquad
+
\frac{1}{4}|[b,C]|^2
+
\frac{1}{2}|[b,\phi_+]|^2
+
\frac{1}{2}|[\bar{b},\phi_+]|^2
+
\frac{1}{4}|[B_3,C]|^2
+
|[B_3,\phi_+]|^2
\Bigl\}, 
\end{eqnarray}

\begin{eqnarray}
\Delta S^{bos}
=
\frac{1}{2g_{0d}^2}Tr\Bigl\{
-\frac{\mu}{2}C[\phi_+,\phi_-]
+
\frac{\mu^2}{9}\left(
\frac{C^2}{4}+\phi_+ \phi_-
-
\frac{2i\mu}{3}B_3[b,\bar{b}]
\right)
\Bigl\}. 
\end{eqnarray}

After projection it reduces to 
\begin{eqnarray}
S_{0}^{bos,lat}
&=&
\frac{1}{2g_{0d}^2}\sum_{\vec{n}}Tr\Bigl\{
\frac{1}{4}|
x_{\vec{n}}\bar{x}_{\vec{n}}
-\bar{x}_{\vec{n}-\hat{x}}x_{\vec{n}-\hat{x}}
+
y_{\vec{n}}\bar{y}_{\vec{n}}
-\bar{y}_{\vec{n}-\hat{y}}y_{\vec{n}-\hat{y}}
|^2
+
|
x_{\vec{n}}y_{\vec{n}+\hat{x}}
-
y_{\vec{n}}x_{\vec{n}+\hat{y}}
|^2
\nonumber\\
& &
+
\frac{1}{4}|[\phi_{+,\vec{n}},\phi_{-,\vec{n}}]|^2
+
\frac{1}{4}|[\phi_{+,\vec{n}},C_{\vec{n}}]|^2
+
\frac{1}{4}|
b_{\vec{n}}\bar{b}_{\vec{n}}
-
\bar{b}_{\vec{n}+\hat{x}-\hat{y}}b_{\vec{n}+\hat{x}-\hat{y}}
|^2
+
|
b_{\vec{n}} B_{3,\vec{n}-\hat{x}+\hat{y}}
-
B_{3,\vec{n}} b_{\vec{n}}
|^2
\nonumber\\
& & 
+
\frac{1}{2}|
x_{\vec{n}}\phi_{+,\vec{n}+\hat{x}}
-\phi_{+,\vec{n}} x_{\vec{n}}
|^2
+
\frac{1}{2}|
\bar{x}_{\vec{n}-\hat{x}}\phi_{+,\vec{n}-\hat{x}}
-
\phi_{+,\vec{n}} \bar{x}_{\vec{n}-\hat{x}}
|^2
+
\frac{1}{4}|
x_{\vec{n}}C_{\vec{n}+\hat{x}}
-
C_{\vec{n}}x_{\vec{n}}
|^2
\nonumber\\
& &
+
\frac{1}{2}|
y_{\vec{n}}\phi_{+,\vec{n}+\hat{y}}
-\phi_{+,\vec{n}} y_{\vec{n}}
|^2
+
\frac{1}{2}|
\bar{y}_{\vec{n}-\hat{y}}\phi_{+,\vec{n}-\hat{y}}
-
\phi_{+,\vec{n}} \bar{y}_{\vec{n}-\hat{y}}
|^2
+
\frac{1}{4}|
y_{\vec{n}}C_{\vec{n}+\hat{y}}
-
C_{\vec{n}}y_{\vec{n}}
|^2
\nonumber\\
& &
+
\frac{1}{2}|
x_{\vec{n}}b_{\vec{n}+\hat{x}}
-
b_{\vec{n}}x_{\vec{n}-\hat{x}+\hat{y}}
|^2
+
\frac{1}{2}|
\bar{x}_{\vec{n}}b_{\vec{n}}
-
b_{\vec{n}+\hat{x}}\bar{x}_{\vec{n}-\hat{x}+\hat{y}}
|^2
+
|
x_{\vec{n}} B_{3,\vec{n}+\hat{x}}
-
B_{3,\vec{n}} x_{\vec{n}}
|^2
\nonumber\\
& &
+
\frac{1}{2}|
y_{\vec{n}}b_{\vec{n}+\hat{y}}
-
b_{\vec{n}}y_{\vec{n}-\hat{x}+\hat{y}}
|^2
+
\frac{1}{2}|
\bar{y}_{\vec{n}-\hat{y}}b_{\vec{n}-\hat{y}}
-
b_{\vec{n}}\bar{y}_{\vec{n}-\hat{x}}
|^2
+
|
y_{\vec{n}} B_{3,\vec{n}+\hat{y}}
-
B_{3,\vec{n}} y_{\vec{n}}
|^2 
\nonumber\\
& &
+
\frac{1}{4}|
b_{\vec{n}}C_{\vec{n}-\hat{x}+\hat{y}}
-
C_{\vec{n}}b_{\vec{n}}
|^2
+
\frac{1}{2}|
b_{\vec{n}}\phi_{+,\vec{n}-\hat{x}+\hat{y}}
-
\phi_{+,\vec{n}} b_{\vec{n}}
|^2
+
\frac{1}{2}|
\bar{b}_{\vec{n}} \phi_{+,\vec{n}}
-
\phi_{+,\vec{n}-\hat{x}+\hat{y}} \bar{b}_{\vec{n}}
|^2
\nonumber\\
& &
+
\frac{1}{4}|[B_{3,\vec{n}},C_{\vec{n}}]|^2
+
|[B_{3,\vec{n}},\phi_{+,\vec{n}}]|^2
\Bigl\}
\end{eqnarray}
and
\begin{eqnarray}
\Delta S^{bos}
=
\frac{1}{2g_{0d}^2}\sum_{\vec{n}}Tr\Bigl\{
-\frac{\mu}{2}C_{\vec{n}}[\phi_{+,\vec{n}},\phi_{-,\vec{n}}]
+
\frac{\mu^2}{9}\left(
\frac{C_{\vec{n}}^2}{4}+\phi_{+,\vec{n}} \phi_{-,\vec{n}}
-
\frac{2i\mu}{3}B_{3,\vec{n}}\left(
b_{\vec{n}}\bar{b}_{\vec{n}}
-
\bar{b}_{\vec{n}+\hat{x}-\hat{y}}
b_{\vec{n}+\hat{x}-\hat{y}}
\right)
\right)
\Bigl\}. 
\end{eqnarray}
SUSY transformation on lattice is shown in appendix. 

If we expand this model around $x_{\vec{n}}\bar{x}_{\vec{n}}
=
y_{\vec{n}}\bar{y}_{\vec{n}}=\frac{1}{a^2}\cdot\textbf{1}$, 2d SYM is obtained. 
To stabilize the background , we add soft SUSY breaking terms\footnote{
In principle we can obtain such term keeping two supersymmetries, by adding 
appropriate terms to ${\cal F}$, though the action becomes ugly. 
For simulation, it might be necessary to add mass for $b,\bar{b}$ and $B_3$ as well. 
It can be done keeping two exact supersymmetries by changing the values 
of $a_A$ in \eqref{mass_deformation}; adding soft SUSY breaking mass is also fine. 
In any case 
the following argument for the absence of fine tuning is not modified. 
}. 
Here we add two kinds of mass terms, 
\begin{eqnarray}
\frac{a^2}{8g_{0d}^2}\sum_{\vec{n}}
\left\{
\nu^2_1
Tr\left(
\left|x_{\vec{n}}\bar{x}_{\vec{n}}-\frac{1}{a^2}\right|^2
+
\left|y_{\vec{n}}\bar{y}_{\vec{n}}-\frac{1}{a^2}\right|^2
\right)
+
\nu_2^2
\left(
\left|
\frac{Tr(x_{\vec{n}}\bar{x}_{\vec{n}})}{N}
-
\frac{1}{a^2}
\right|^2
+
\left|
\frac{Tr(y_{\vec{n}}\bar{y}_{\vec{n}})}{N}
-
\frac{1}{a^2}
\right|^2
\right)
\right\}. 
\label{SYM_soft_mass}
\end{eqnarray} 
Two-dimensional coupling constant $g_{2d}$ is given as before, 
\begin{eqnarray}
g_{2d}^2 = a^2 g_{0d}^2. 
\end{eqnarray}
A parameter $\nu_1$ gives mass of both $SU(N)$ and $U(1)$ parts, 
while $\nu_2$ is solely $U(1)$ mass\footnote{
We would like to thank O.~Aharony for his suggestion of 
$U(1)$ mass term. 
}. 
As argued in \cite{AHNT} and \cite{Hanada:2009hq}, 
$SU(N)$ flat direction is lifted by quantum effects 
and nonabelian phase (i.e. a phase 
in which all scalar eigenvalues localizes to a point) 
becomes stable at large-$N$. Therefore, we can send $\nu_1$ to zero 
when we consider the large-$N$ limit.   
On the other hand, to stabilize $U(1)$ flat direction, 
we must take $\nu_2$ to be $O(1)$. 
However it does not affect the supersymmetry in $SU(N)$ sector, 
because in the continuum limit $U(1)$ part is just a decoupled free sector. 

Note that we can introduce $U(1)$ mass for other scalars as well, although 
those $U(1)$ modes are harmless. 
\subsection{Continuum limit and absence of fine tuning}
At tree level, continuum action is\footnote{
Here $X_i$ stand for 
$X_1=(\phi_+ + \phi_-)/2$, 
$X_2=(\phi_+ - \phi_-)/2i$, 
$X_3=C/2$, 
$X_{4,5}=s_{1,2}$, 
$X_{6,7,8}=B_{1,2,3}$. 
} (here we show only bosonic part; Also we omitted $U(1)$ sector.)
\begin{eqnarray}
S^{bos,cont}
&=&
\frac{1}{g_{2d}^2}\int d^2x\ 
Tr\Bigl\{
\frac{1}{2}F_{12}^2
+
\frac{1}{2}(D_\mu X_I)^2
-
\frac{1}{4}[X_I,X_J]^2
+
\frac{\mu^2}{18}\sum_{a=1}^3 X_a^2
+
2i\mu X_1[X_2,X_3]
\nonumber\\ 
& &
\qquad
-
\frac{4\mu}{3} X_6[X_7,X_{8}]
+
\frac{\nu^2_1}{2}\sum_{i=4}^5 X_i^2
\Bigl\}. 
\label{continuum_bosonic}
\end{eqnarray}
When $\nu_1=0$, deformation by $\mu$ breaks 14 out of 16 SUSY softly. 
With nonzero $\nu_1$ remaining two is softly broken as well. 

In the following we show that, at perturbative level, 
radiative corrections from UV region do not change this continuum action, 
except for the soft terms due to nonzero value of $\nu_1$. 
Because our lattice has the same symmetries as the model in \cite{Hanada:2010kt}, 
the argument goes completely in parallel. 
Firstly let us consider the case that $\nu_1=0$. 
Let us consider operators of the form ${\cal O}_p=\varphi^\alpha\partial^\beta\phi^{2\gamma}$, 
whose mass dimension is $p=\alpha+\beta+3\gamma$. Here $\varphi$ and $\phi$ stand for 
boson and fermion, respectively. 
Because the coupling constant $g_{2d}$ is dimensionful, the correction is of the form 
\begin{eqnarray}
\left(
c_1a^{p-2}
+
g_{2d}^2 c_2 a^p
+
\cdots
\right)\int d^2x{\cal O}_p(x). 
\end{eqnarray}
Here $a^{p-4}/g_{2d}^2$ is omitted because it is a tree-level contribution. 
Therefore, nonzero contribution may remain in the continuum limit $a\to 0$ only when $p=1,2$, 
and if that happens ${\cal O}$ must be $\varphi$ or $\varphi^2$. 
But all such terms are forbidden by exact $Q_\pm$ SUSY and $SU(2)$ R-symmetry.  
Even if $\nu_1$ is not zero, it is just a soft mass, which does not alter 
UV divergence. Therefore in an appropriate large-volume and $\nu_1\to 0$ limit 
SUSY breaking effect disappears \cite{Cohen:2003xe,KaplanUnsal16SUSY}. 
When we consider an uplift to 4d theory in the following, $N$ is sent to infinity, 
and then scalar flat direction is lifted by the quantum effect. 
Therefore we can take $\nu_1$ to be zero. 
\subsection{Remarks}\label{sec:remarks}
The argument provided in this section applies to all order in perturbation theory. 
Whether the absence of the sign problem persists to nonperturbative level should be checked 
by numerics. In four-SUSY cousin of this model the absence of the fine tuning at nonperturbative 
level has been confirmed in \cite{Hanada:2009hq,Hanada:2010qg} for two independent formulations 
by Sugino and by Cohen-Kaplan-Katz-Unsal (the latter corresponds to the one adopted in this paper), 
by treating fermions dynamically.  
It is natural to expect that the absence of fine tuning persists to nonperturbative level 
also in the present case with maximal supersymmetry. 

However, with maximal supersymmetry, there is a possible difficulty 
for the Monte-Carlo simulation --  
Pfaffian of the Dirac operator is complex in general and hence 
usual Markov chain Monte-Carlo method cannot be used. 
At finite temperature, the phase is almost absent even at rather low temperature 
which is physically interesting, and phase quench approximation, in which 
the Pfaffian is replaced by its absolute value, gives a good approximation 
\cite{AHNT,Catterall-Wiseman,Catterall:2010fx}. (Strictly speaking, the vacuum 
studied in these works are different from the fuzzy sphere. Also, in order to obtain 
4d theory one has to take a certain scaling limit. Therefore, whether the phase is 
small in the present case must be checked independently.)   
It enable us to test the validity of the model at nonperturbative level. 
Surprisingly, in one dimension, 
even at very low temperature and/or with SUSY-preserving boundary condition, 
where the phase fluctuates, the phase quench approximation provide a correct results 
predicted by the gauge/gravity duality \cite{AHNT}\footnote{Numerically it can be justified by looking 
at the correlation between the phase and values of observables. }. 
It is interesting to check whether the same happens in the present case. 

\section{Uplift to 4d}\label{sec:4d}
For the reason explained above, we assume $\nu_1=0$. 
The continuum action \eqref{continuum_bosonic} has constant BPS fuzzy sphere solution\footnote{
This background preserves exact supersymmetries at discretized level. 
} 
\begin{eqnarray}
X_a(x)=\frac{\mu}{3}L_a
\qquad(a=1,2,3), 
\qquad
X_i(x)=0
\qquad(i=4,\cdots,8), 
\end{eqnarray}
where $L_a$ are $M\times M$ matrices satisfying $SU(2)$ commutation relation 
\begin{eqnarray}
 [L_a,L_b]=i\epsilon_{abc}L_c. 
\end{eqnarray} 
By taking $k$-coincident fuzzy sphere solution, 
$L_a=L_a^{(M/k)}\otimes\textbf{1}_k$, where $L_a^{(M/k)}$ is the $(M/k)\times (M/k)$ 
irreducible representation, we obtain 4d $U(k)$ theory on fuzzy sphere.  
Essentially, adjoint action of $L_a$ is identified with the derivative 
and $[X_a,\ \cdot\ ]$ is regarded as the gauge covariant derivative 
\cite{Aoki:1999vr}. 
The noncommutativity is given by $\theta\sim k/(\mu^2 M)$ and UV/IR momentum cutoffs 
along spherical directions are $\mu M/k$ and $\mu$, respectively. 
4d coupling is given by $g_{4d}^2 = 4\pi\theta g_{2d}^2$. 
In order to get continuum 4d theory, we take large-$M$ and small $\mu$ limit while 
fixing $k$ and $g_{4d}^2$. 
In that limit, maximal supersymmetry is restored 
because soft SUSY breaking parameter $\mu$ goes to zero. 
One can take a limit with any value of noncommutativity $\theta$, 
and $\theta\to 0$ limit is expected to be smooth \cite{Hashimoto:1999ut,Matusis:2000jf}. 
That the limit should be smooth is natural physically, because a possible obstacle 
is a new IR divergence arising due to the UV/IR mixing reflecting the UV divergence, 
which should be absent in UV finite theories. 
However there is no rigorous proof in mathematical sense. 
Our formulation itself can serve as a nonperturbative framework to check the smoothness.  

In the above we assumed the radius of the fuzzy sphere does not deviate from classical value. 
Whether it is the case or not should be tested by numerics. If the radius is renormalized, 
we should take into account it by replacing the parameters in the mapping rule with 
renormalized ones. 

\section{Conclusion and discussions}
In this paper we proposed a nonperturbative regularization of 4d ${\cal N}=4$ SYM 
which does not require parameter fine tuning at least at perturbative level. 
It is much simpler than similar model \cite{Hanada:2010kt} and can easily be put on computer.  
Therefore absence of the fine tuning at nonperturbative level can be tested more easily.  
Except for the lack of a mathematical proof of the smoothness 
of the commutative limit of the noncommutative space, 
which can be tested numerically by using this model itself, this method provides the first formulation 
of four-dimensional extended SYM free from the fine tuning at perturbative level. 
Note that in other models absence of fine tuning is not shown even at perturbative level. 
We expect lattice Monte-Carlo simulation will be performed in near future and 
new insights into $AdS_5/CFT_4$ correspondence will be obtained. 

Soft mass term for $U(1)$ scalar introduced in \eqref{SYM_soft_mass} is very simple 
but can play an important role. Consider the original, non-deformed model. 
As already mentioned, $SU(N)$ flat direction is lifted at large-$N$, and hence 
$U(N)$ mass term $\nu_1$ can be set to zero. Therefore, even at finite volume, 
maximal supersymmetry is fully restored. Such a finite volume theory has 
a gravity dual description \cite{Aharony:2004ig} of the black hole/black string transition 
\cite{Gregory:1993vy}. This theory is expected to have a rich phase structure 
as a function of volume and temperature \cite{Aharony:2004ig,Kawahara:2007fn}, 
and details of the geometry of the transition can be studied 
from Monte-Carlo data \cite{Azeyanagi:2009zf}. Hence Monte-Carlo simulation 
would provide valuable insights into the stringy correction to the transition\footnote{
For recent simulations in this context, see \cite{Hanada:2009hq,Catterall:2010fx}. 
}.


\section*{Acknowledgments}
We would like to thank O.~Aharony, I.~Kanamori, S.~Matsuura 
and M.~\"{U}nsal  for stimulating 
discussions and comments.

\appendix
\section{Fermionic part of the lattice action}
Fermionic part of the two-dimensional lattice action is written as 
\begin{eqnarray}
S^{fer}=S^{fer}_0+\Delta S^{fer}, 
\qquad
S^{fer}_0
=
\sum_{i=1}^{10}
\frac{1}{2g_{0d}^2}\sum_{\vec{n}} {\cal L}_i, 
\end{eqnarray}
where



\begin{eqnarray}
{\cal L}_1
&=&
-\bar{\rho}_{+,\vec{n}}\left(
(
\bar{\xi}_{-x,\vec{n}-\hat{x}}y_{\vec{n}-\hat{x}}
-
y_{\vec{n}}\bar{\xi}_{-x,\vec{n}-\hat{x}+\hat{y}}
)
+
(
\bar{x}_{\vec{n}-\hat{x}}\xi_{-y,\vec{n}-\hat{x}}
-
\xi_{-y,\vec{n}}\bar{x}_{\vec{n}-\hat{x}+\hat{y}}
)
\right)
\nonumber\\
& &
+
\bar{\rho}_{-,\vec{n}}\left(
(
\bar{\xi}_{+x,\vec{n}-\hat{x}}y_{\vec{n}-\hat{x}}
-
y_{\vec{n}}\bar{\xi}_{+x,\vec{n}-\hat{x}+\hat{y}}
)
+
(
\bar{x}_{\vec{n}-\hat{x}}\xi_{+y,\vec{n}-\hat{x}}
-
\xi_{+y,\vec{n}}\bar{x}_{\vec{n}-\hat{x}+\hat{y}}
)
\right)
\nonumber\\
& & 
-
\bar{b}_{\vec{n}}
\left(
(
\bar{\xi}_{+x,\vec{n}-\hat{x}}\xi_{-y,\vec{n}-\hat{x}}
+
\xi_{-y,\vec{n}}\bar{\xi}_{+x,\vec{n}-\hat{x}+\hat{y}}
)
-
(
\bar{\xi}_{-x,\vec{n}-\hat{x}}\xi_{+y,\vec{n}-\hat{x}}
+
\xi_{+y,\vec{n}}\bar{\xi}_{-x,\vec{n}-\hat{x}+\hat{y}}
)
\right), 
\end{eqnarray}


\begin{eqnarray}
{\cal L}_2
&=&
-\rho_{+,\vec{n}+\hat{x}-\hat{y}}\left(
(
\xi_{-x,\vec{n}}\bar{y}_{\vec{n}+\hat{x}-\hat{y}}
-
\bar{y}_{\vec{n}-\hat{y}}\xi_{-x,\vec{n}-\hat{y}}
)
+
(
x_{\vec{n}}\bar{\xi}_{-y,\vec{n}+\hat{x}-\hat{y}}
-
\bar{\xi}_{-y,\vec{n}-\hat{y}}x_{\vec{n}-\hat{y}}
)
\right)
\nonumber\\
& &
+
\rho_{-,\vec{n}+\hat{x}-\hat{y}}\left(
(
\xi_{+x,\vec{n}}\bar{y}_{\vec{n}+\hat{x}-\hat{y}}
-
\bar{y}_{\vec{n}-\hat{y}}\xi_{+x,\vec{n}-\hat{y}}
)
+
(
x_{\vec{n}}\bar{\xi}_{+y,\vec{n}+\hat{x}-\hat{y}}
-
\bar{\xi}_{+y,\vec{n}-\hat{y}}x_{\vec{n}-\hat{y}}
)
\right)
\nonumber\\ 
& &
-
b_{\vec{n}+\hat{x}-\hat{y}}
\left(
(
\xi_{+x,\vec{n}}\bar{\xi}_{-y,\vec{n}+\hat{x}-\hat{y}}
+
\bar{\xi}_{-y,\vec{n}-\hat{y}}\xi_{+x,\vec{n}-\hat{y}}
)
-
(
\xi_{-x,\vec{n}}\bar{\xi}_{+y,\vec{n}+\hat{x}-\hat{y}}
+
\bar{\xi}_{+y,\vec{n}-\hat{y}}\xi_{-x,\vec{n}-\hat{y}}
)
\right),  
\nonumber\\
\end{eqnarray}


\begin{eqnarray}
{\cal L}_3
&=& 
-i\chi_{+3,\vec{n}}
\left(
(
\xi_{-x,\vec{n}}\bar{x}_{\vec{n}}
-
\bar{x}_{\vec{n}-\hat{x}}\xi_{-x,\vec{n}-\hat{x}}
)
+
(
x_{\vec{n}}\bar{\xi}_{-x,\vec{n}}
-
\bar{\xi}_{-x,\vec{n}-\hat{x}}x_{\vec{n}-\hat{x}}
)
\right) 
\nonumber\\
& &
+i\chi_{-3,\vec{n}}
\left(
(
\xi_{+x,\vec{n}}\bar{x}_{\vec{n}}
-
\bar{x}_{\vec{n}-\hat{x}}\xi_{+x,\vec{n}-\hat{x}}
)
+
(
x_{\vec{n}}\bar{\xi}_{+x,\vec{n}}
-
\bar{\xi}_{+x,\vec{n}-\hat{x}}x_{\vec{n}-\hat{x}}
)
\right) 
\nonumber\\
& &
-
iB_{3,\vec{n}}
\left(
(
\xi_{+x,\vec{n}}\bar{\xi}_{-x,\vec{n}}
+
\bar{\xi}_{-x,\vec{n}-\hat{x}}\xi_{+x,\vec{n}-\hat{x}}
)
-
(
\xi_{-x,\vec{n}}\bar{\xi}_{+x,\vec{n}}
+
\bar{\xi}_{+x,\vec{n}-\hat{x}}\xi_{-x,\vec{n}-\hat{x}}
)
\right), 
\end{eqnarray}


\begin{eqnarray}
{\cal L}_4
&=& 
i\chi_{+3,\vec{n}}
\left(
(
\xi_{-y,\vec{n}}\bar{y}_{\vec{n}}
-
\bar{y}_{\vec{n}-\hat{y}}\xi_{-y,\vec{n}-\hat{y}}
)
+
(
y_{\vec{n}}\bar{\xi}_{-y,\vec{n}}
-
\bar{\xi}_{-y,\vec{n}-\hat{y}}y_{\vec{n}-\hat{y}}
)
\right) 
\nonumber\\
& &
-i\chi_{-3,\vec{n}}
\left(
(
\xi_{+y,\vec{n}}\bar{y}_{\vec{n}}
-
\bar{y}_{\vec{n}-\hat{y}}\xi_{+y,\vec{n}-\hat{y}}
)
+
(
y_{\vec{n}}\bar{\xi}_{+y,\vec{n}}
-
\bar{\xi}_{+y,\vec{n}-\hat{y}}y_{\vec{n}-\hat{y}}
)
\right) 
\nonumber\\
& &
+
iB_{3,\vec{n}}
\left(
(
\xi_{+y,\vec{n}}\bar{\xi}_{-y,\vec{n}}
+
\bar{\xi}_{-y,\vec{n}-\hat{y}}\xi_{+y,\vec{n}-\hat{y}}
)
-
(
\xi_{-y,\vec{n}}\bar{\xi}_{+y,\vec{n}}
+
\bar{\xi}_{+y,\vec{n}-\hat{y}}\xi_{-y,\vec{n}-\hat{y}}
)
\right), 
\end{eqnarray}


\begin{eqnarray}
{\cal L}_5
&=& 
-i\chi_{+3,\vec{n}}\left(
(
\rho_{-,\vec{n}}\bar{b}_{\vec{n}}
-
\bar{b}_{\vec{n}+\hat{x}-\hat{y}}\rho_{-,\vec{n}+\hat{x}-\hat{y}}
)
+
(
b_{\vec{n}}\bar{\rho}_{-,\vec{n}}
-
\bar{\rho}_{-,\vec{n}+\hat{x}-\hat{y}}b_{\vec{n}+\hat{x}-\hat{y}}
)
\right)
\nonumber\\
& &
+
i\chi_{-3,\vec{n}}\left(
(
\rho_{+,\vec{n}}\bar{b}_{\vec{n}}
-
\bar{b}_{\vec{n}+\hat{x}-\hat{y}}\rho_{+,\vec{n}+\hat{x}-\hat{y}}
)
+
(
b_{\vec{n}}\bar{\rho}_{+,\vec{n}}
-
\bar{\rho}_{+,\vec{n}+\hat{x}-\hat{y}}b_{\vec{n}+\hat{x}-\hat{y}}
)
\right)
\nonumber\\
& &
-iB_{3,\vec{n}}\left(
(
\rho_{+,\vec{n}}\bar{\rho}_{-,\vec{n}}
+
\bar{\rho}_{-,\vec{n}+\hat{x}-\hat{y}}\rho_{+,\vec{n}+\hat{x}-\hat{y}}
)
-
(
\rho_{-,\vec{n}}\bar{\rho}_{+,\vec{n}}
+
\bar{\rho}_{+,\vec{n}+\hat{x}-\hat{y}}\rho_{-,\vec{n}+\hat{x}-\hat{y}}
)
\right), 
\end{eqnarray}


\begin{eqnarray}
{\cal L}_6
&=&
-\sum_{\vec{n}}
\frac{1}{2}
\Bigl\{
2\bar{\xi}_{+x,\vec{n}}
(
\xi_{+x,\vec{n}}\phi_{-,\vec{n}+\hat{x}}
-
\phi_{-,\vec{n}}\xi_{+x,\vec{n}}
)
\nonumber\\
& &
+
\bar{\xi}_{+x,\vec{n}}
(
x_{\vec{n}}\eta_{-,\vec{n}+\hat{x}}
-
\eta_{-,\vec{n}}x_{\vec{n}}
)
+
\xi_{+x,\vec{n}}
(
\bar{x}_{\vec{n}}\eta_{-,\vec{n}}
-
\eta_{-,\vec{n}+\hat{x}}\bar{x}_{\vec{n}}
)
\nonumber\\
& &
+
\left(
(
C_{\vec{n}+\hat{x}}\bar{\xi}_{+x,\vec{n}}
-
\bar{\xi}_{+x,\vec{n}}C_{\vec{n}}
)
-
(
\bar{x}_{\vec{n}}\eta_{+,\vec{n}}
-
\eta_{+,\vec{n}+\hat{x}}\bar{x}_{\vec{n}}
)
-
(
\phi_{+,\vec{n}+\hat{x}}\bar{\xi}_{-x,\vec{n}}
-
\bar{\xi}_{-x,\vec{n}}\phi_{+,\vec{n}}
)
\right)\xi_{-x,\vec{n}}
\nonumber\\
& &
+
\left(
(
C_{\vec{n}}\xi_{+x,\vec{n}}
-
\xi_{+x,\vec{n}}C_{\vec{n}+\hat{x}}
)
-
(
x_{\vec{n}}\eta_{+,\vec{n}+\hat{x}}
-
\eta_{+,\vec{n}}x_{\vec{n}}
)
-
(
\phi_{+,\vec{n}}\xi_{-x,\vec{n}}
-
\xi_{-x,\vec{n}}\phi_{+,\vec{n}+\hat{x}}
) 
\right)\bar{\xi}_{-x,\vec{n}}
\Bigl\},   
\nonumber\\
\end{eqnarray}


\begin{eqnarray}
{\cal L}_7
&=&
-\sum_{\vec{n}}
\frac{1}{2}
\Bigl\{
2\bar{\xi}_{+y,\vec{n}}
(
\xi_{+y,\vec{n}}\phi_{-,\vec{n}+\hat{y}}
-
\phi_{-,\vec{n}}\xi_{+y,\vec{n}}
)
\nonumber\\
& &
+
\bar{\xi}_{+y,\vec{n}}
(
y_{\vec{n}}\eta_{-,\vec{n}+\hat{y}}
-
\eta_{-,\vec{n}}y_{\vec{n}}
)
+
\xi_{+y,\vec{n}}
(
\bar{y}_{\vec{n}}\eta_{-,\vec{n}}
-
\eta_{-,\vec{n}+\hat{y}}\bar{y}_{\vec{n}}
)
\nonumber\\
& &
+
\left(
(
C_{\vec{n}+\hat{y}}\bar{\xi}_{+y,\vec{n}}
-
\bar{\xi}_{+y,\vec{n}}C_{\vec{n}}
)
-
(
\bar{y}_{\vec{n}}\eta_{+,\vec{n}}
-
\eta_{+,\vec{n}+\hat{y}}\bar{y}_{\vec{n}}
)
-
(
\phi_{+,\vec{n}+\hat{y}}\bar{\xi}_{-y,\vec{n}}
-
\bar{\xi}_{-y,\vec{n}}\phi_{+,\vec{n}}
)
\right)\xi_{-y,\vec{n}}
\nonumber\\
& &
+
\left(
(
C_{\vec{n}}\xi_{+y,\vec{n}}
-
\xi_{+y,\vec{n}}C_{\vec{n}+\hat{y}}
)
-
(
y_{\vec{n}}\eta_{+,\vec{n}+\hat{y}}
-
\eta_{+,\vec{n}}y_{\vec{n}}
)
-
(
\phi_{+,\vec{n}}\xi_{-y,\vec{n}}
-
\xi_{-y,\vec{n}}\phi_{+,\vec{n}+\hat{y}}
) 
\right)\bar{\xi}_{-y,\vec{n}}
\Bigl\},   
\nonumber\\
\end{eqnarray}


\begin{eqnarray}
{\cal L}_8
&=&  
-\sum_{\vec{n}}
\frac{1}{2}
\Bigl\{
2\bar{\rho}_{+,\vec{n}}(
\rho_{+,\vec{n}}\phi_{-,\vec{n}-\hat{x}+\hat{y}}
-
\phi_{-,\vec{n}}\rho_{-,\vec{n}}
)
\nonumber\\
& &
+
\bar{\rho}_{+,\vec{n}}(
b_{\vec{n}}\eta_{-,\vec{n}-\hat{x}+\hat{y}}
-
\eta_{-,\vec{n}}b_{\vec{n}}
)
+
\rho_{+,\vec{n}}(
\bar{b}_{\vec{n}}\eta_{-,\vec{n}}
-
\eta_{-,\vec{n}-\hat{x}+\hat{y}}\bar{b}_{\vec{n}}
)
\nonumber\\
& &
+
\left(
(
C_{\vec{n}-\hat{x}+\hat{y}}\bar{\rho}_{+,\vec{n}}
-
\bar{\rho}_{+,\vec{n}}C_{\vec{n}}
)
-
(\bar{b}_{\vec{n}}\eta_{+,\vec{n}}
-
\eta_{+,\vec{n}-\hat{x}+\hat{y}}\bar{b}_{\vec{n}}
)
-
(
\phi_{+,\vec{n}-\hat{x}+\hat{y}}\bar{\rho}_{-,\vec{n}}
-
\bar{\rho}_{-,\vec{n}}\phi_{+,\vec{n}}
) 
\right)\rho_{-,\vec{n}}
\nonumber\\
& &
+
\left(
(
C_{\vec{n}}\rho_{+,\vec{n}}
-
\rho_{+,\vec{n}}C_{\vec{n}-\hat{x}+\hat{y}}
)
-
(
b_{\vec{n}}\eta_{+,\vec{n}-\hat{x}+\hat{y}}
-
\eta_{+,\vec{n}}b_{\vec{n}}
)
-
(
\phi_{+,\vec{n}}\rho_{-,\vec{n}}
-
\rho_{-,\vec{n}}\phi_{+,\vec{n}-\hat{x}+\hat{y}}
)
\right)\bar{\rho}_{-,\vec{n}}
\Bigl\},   
\nonumber\\
\end{eqnarray}


\begin{eqnarray}
{\cal L}_9
&=&  
\sum_{\vec{n}}\Bigl\{
\chi_{+3,\vec{n}}[\eta_{-,\vec{n}},B_{3,\vec{n}}]
+
\chi_{-3,\vec{n}}[\eta_{+,\vec{n}},B_{3,\vec{n}}]
\nonumber\\
& &
+
\chi_{+3,\vec{n}}[\phi_{-,\vec{n}},\chi_{+3,\vec{n}}]
-
\chi_{-3,\vec{n}}[\phi_{+,\vec{n}},\chi_{-3,\vec{n}}]
+
[\chi_{+3,\vec{n}},C_{\vec{n}}]\chi_{-3,\vec{n}}
\Bigl\}, 
\nonumber\\
\end{eqnarray}


\begin{eqnarray}
{\cal L}_{10}
&=&  
\sum_{\vec{n}}
\frac{1}{4}\left\{
\eta_{+,\vec{n}}[\eta_{-,\vec{n}},C_{\vec{n}}]
+
\eta_{+,\vec{n}}[\phi_{-,\vec{n}},\eta_{+,\vec{n}}]
-
\eta_{-,\vec{n}}[\phi_{+,\vec{n}},\eta_{-,\vec{n}}]
\right\}
\end{eqnarray}
and 
\begin{eqnarray}
\Delta S^{fer}
=
\frac{\mu}{2g_{0d}^2}\sum_{\vec{n}}Tr
\left\{
\frac{1}{3}
\left(
\bar{\xi}_{+x,\vec{n}}\xi_{-x,\vec{n}}
+
\xi_{+x,\vec{n}}\bar{\xi}_{-x,\vec{n}}
+
\bar{\xi}_{+y,\vec{n}}\xi_{-y,\vec{n}}
+
\xi_{+y,\vec{n}}\bar{\xi}_{-y,\vec{n}}
\right)
-
\frac{1}{6}
\eta_{+,\vec{n}}\eta_{-,\vec{n}}
\right\}. 
\end{eqnarray}

\section{SUSY transformation on lattice}

SUSY transformation on two-dimensional lattice is given by 
\begin{eqnarray}
& &
Q^{(0)}_{\pm}x_{\vec{n}}=\xi_{\pm x, \vec{n}},
\quad
Q^{(0)}_\pm\xi_{\pm x, \vec{n}}
=
\mp (x_{\vec{n}}\phi_{\pm, \vec{n}+\hat{x}}
-
\phi_{\pm, \vec{n}}x_{\vec{n}}), 
\nonumber\\
& &
Q^{(0)}_\mp\xi_{\pm x,\vec{n}}
=
-\frac{1}{2}(x_{\vec{n}} C_{\vec{n}+\hat{x}} - C_{\vec{n}}x_{\vec{n}})
\mp\tilde{h}_{x,\vec{n}}, 
\nonumber\\
& &
Q^{(0)}_{\pm}y_{\vec{n}}=\xi_{\pm y,\vec{n}},
\quad
Q^{(0)}_\pm\xi_{\pm y,\vec{n}}
=
\mp \left(
y_{\vec{n}}\phi_{\pm,\vec{n}+\hat{y}}
-
\phi_{\pm,\vec{n}} y_{\vec{n}}
\right), 
\nonumber\\
& &
Q^{(0)}_\mp\xi_{\pm y, \vec{n}}
=
-\frac{1}{2}(
y_{\vec{n}} C_{\vec{n}+\hat{y}}
-
C_{\vec{n}}y_{\vec{n}}
)
\mp\tilde{h}_{y,\vec{n}}, 
\nonumber\\
& &
Q^{(0)}_{\pm}\tilde{h}_{x,\vec{n}}
=
\left(
\phi_{\pm,\vec{n}}\xi_{\mp x,\vec{n}}
-
\xi_{\mp x,\vec{n}}\phi_{\pm,\vec{n}+\hat{x}}
\right)
\mp\frac{1}{2}\left(
C_{\vec{n}}\xi_{\pm x,\vec{n}}
-
\xi_{\pm x,\vec{n}} C_{\vec{n}+\hat{x}}
\right)
\pm\frac{1}{2}\left(
x_{\vec{n}}\eta_{\pm,\vec{n}+\hat{x}}
-
\eta_{\pm,\vec{n}} x_{\vec{n}}
\right), 
\nonumber\\
& &
Q^{(0)}_{\pm}\tilde{h}_{y,\vec{n}}
=
\left(
\phi_{\pm,\vec{n}}\xi_{\mp y,\vec{n}}
-
\xi_{\mp y,\vec{n}}\phi_{\pm,\vec{n}+\hat{y}}
\right)
\mp\frac{1}{2}\left(
C_{\vec{n}}\xi_{\pm y,\vec{n}}
-
\xi_{\pm y,\vec{n}} C_{\vec{n}+\hat{y}}
\right)
\pm\frac{1}{2}\left(
y_{\vec{n}}\eta_{\pm,\vec{n}+\hat{y}}
-
\eta_{\pm,\vec{n}} y_{\vec{n}}
\right), 
\nonumber\\
& &
Q^{(0)}_{\pm}b_{\vec{n}}
=
\rho_{\pm,\vec{n}}, 
\quad
Q^{(0)}_{\pm}\rho_{\pm,\vec{n}}
=
\pm(
\phi_{\pm,\vec{n}} b_{\vec{n}}
-
b_{\vec{n}}\phi_{\pm,\vec{n}-\hat{x}+\hat{y}}
), 
\nonumber\\
& &
Q^{(0)}_{\pm}B_{3,\vec{n}}
=
\chi_{\pm 3,\vec{n}}, 
\quad
Q^{(0)}_{\pm}\chi_{\pm 3,\vec{n}}
=
\pm[\phi_{\pm,\vec{n}},B_{3,\vec{n}}], 
\nonumber\\
& &
Q^{(0)}_\mp\rho_{\pm,\vec{n} }
=
-\frac{1}{2}(
b_{\vec{n}}C_{\vec{n}-\hat{x}+\hat{y}}
-
C_{\vec{n}}b_{\vec{n}}
)
\mp h_{\vec{n}} 
\nonumber\\
& &
Q^{(0)}_\mp\chi_{\pm 3,\vec{n}}
=
-\frac{1}{2}[B_{3,\vec{n}},C_{\vec{n}}]
\mp H_{3,\vec{n}}, 
\nonumber\\
& &
Q^{(0)}_\pm h_{\vec{n}}
=
(
\phi_{\pm,\vec{n}}\rho_{\mp,\vec{n}}
-
\rho_{\mp,\vec{n}}\phi_{\pm,\vec{n}-\hat{x}+\hat{y}}
)
\pm\frac{1}{2}\left(
b_{\vec{n}}\eta_{\pm,\vec{n}-\hat{x}+\hat{y}}
-
\eta_{\pm,\vec{n}} b_{\vec{n}}
\right)
\mp\frac{1}{2}\left(
C_{\vec{n}}\rho_{\pm,\vec{n}}
-
\rho_{\pm \vec{n}} C_{\vec{n}-\hat{x}+\hat{y}}
\right), 
\nonumber\\
& &
Q^{(0)}_\pm H_{3,\vec{n}}
=
[\phi_{\pm,\vec{n}},\chi_{\mp 3,\vec{n}}]
\pm\frac{1}{2}\left[B_{3,\vec{n}},\eta_{\pm,\vec{n}}\right]
\mp\frac{1}{2}\left[C_{\vec{n}},\chi_{\pm 3,\vec{n}}\right], 
\nonumber\\ 
& &
Q^{(0)}_{\pm} C_{\vec{n}}
=
\eta_{\pm,\vec{n}}, 
\quad
Q^{(0)}_\pm\eta_{\pm,\vec{n}}
=
\pm\left[\phi_{\pm,\vec{n}},C_{\vec{n}}\right], 
\nonumber\\
& &
Q^{(0)}_\mp\eta_{\pm,\vec{n}}
=
\mp\left[\phi_{+,\vec{n}},\phi_{-,\vec{n}}\right], 
\nonumber\\
& &
Q^{(0)}_\pm\phi_{\pm,\vec{n}}=0, 
\quad
Q^{(0)}_{\mp}\phi_{\pm,\vec{n}}=\mp\eta_{\pm,\vec{n}}
\end{eqnarray} 
and
\begin{eqnarray}
& & 
\Delta Q_{\pm}\tilde{h}_{x, \vec{n}}
= 
\frac{\mu}{3}\xi_{\pm x, \vec{n}}, 
\qquad
\Delta Q_{\pm}\tilde{h}_{y, \vec{n}}
= 
\frac{\mu}{3}\xi_{\pm y,\vec{n}}, 
\qquad
\Delta Q_\pm h
= 
\frac{\mu}{3}\rho_{\pm, \vec{n}}, 
\nonumber\\
& & 
\Delta Q_\pm H_{3, \vec{n}}
= 
\frac{\mu}{3}\chi_{\pm 3, \vec{n}}, 
\qquad
\Delta Q_\pm\eta_{\pm, \vec{n}}
= 
\frac{2\mu}{3}\phi_{\pm, \vec{n}}, 
\qquad
\Delta Q_\mp\eta_{\pm, \vec{n}}
= 
\pm\frac{\mu}{3}C_{\vec{n}}. 
\end{eqnarray}




\begin{thebibliography}{99}



\bibitem{BFSS}
T.~Banks, W.~Fischler, S.~H.~Shenker and L.~Susskind,
``M theory as a matrix model: A conjecture,''
Phys.\ Rev.\ D {\bf 55}, 5112 (1997),    
[arXiv:hep-th/9610043].
\bibitem{IKKT}
N.~Ishibashi, H.~Kawai, Y.~Kitazawa and A.~Tsuchiya,
``A large-N reduced model as superstring,''
Nucl.\ Phys.\ B {\bf 498}, 467 (1997),  
[arXiv:hep-th/9612115].
%
\bibitem{MatrixString}
  L.~Motl,
  ``Proposals on nonperturbative superstring interactions,''
  arXiv:hep-th/9701025; 
 R.~Dijkgraaf, E.~P.~Verlinde and H.~L.~Verlinde,
  ``Matrix string theory,''
  Nucl.\ Phys.\  B {\bf 500}, 43 (1997),  
  [arXiv:hep-th/9703030].

\bibitem{Maldacena:1997re}
  J.~M.~Maldacena, 
{\it The large N limit of superconformal field theories and supergravity},
Adv.\ Theor.\ Math.\ Phys.\  {\bf 2}, 231 (1998); 
  [Int.\ J.\ Theor.\ Phys.\  {\bf 38} (1999) 1113]
  [arXiv:hep-th/9711200]. 
%
  N.~Itzhaki, J.~M.~Maldacena, J.~Sonnenschein and S.~Yankielowicz,
\emph{Supergravity and the large N limit of theories with sixteen
  supercharges},
Phys.\ Rev.\ D {\bf 58}, 046004 (1998).

\bibitem{Giedt:2008xm}
  J.~Giedt, R.~Brower, S.~Catterall, G.~T.~Fleming and P.~Vranas,
  ``Lattice super-Yang-Mills using domain wall fermions in the chiral limit,''
  Phys.\ Rev.\  D {\bf 79}, 025015 (2009)
  [arXiv:0810.5746 [hep-lat]].


  M.~G.~Endres,
  ``Dynamical simulation of N=1 supersymmetric Yang-Mills theory with domain
  wall fermions,''
  Phys.\ Rev.\  D {\bf 79}, 094503 (2009)
  [arXiv:0902.4267 [hep-lat]].


  K.~Demmouche, F.~Farchioni, A.~Ferling, I.~Montvay, G.~Munster, E.~E.~Scholz and J.~Wuilloud,
  ``Simulation of 4d N=1 supersymmetric Yang-Mills theory with Symanzik
  improved gauge action and stout smearing,''
  arXiv:1003.2073 [hep-lat].


\bibitem{Catterall:2009it}
  S.~Catterall, D.~B.~Kaplan and M.~Unsal,
  ``Exact lattice supersymmetry,''
  Phys.\ Rept.\  {\bf 484}, 71 (2009)
  [arXiv:0903.4881 [hep-lat]].

\bibitem{Hanada:2010qg}
  M.~Hanada and I.~Kanamori,
  ``Absence of sign problem in two-dimensional N=(2,2) super Yang-Mills on
  lattice,''
  arXiv:1010.2948 [hep-lat].


\bibitem{Hanada-Nishimura-Takeuchi}
  M.~Hanada, J.~Nishimura and S.~Takeuchi,
  ``Non-lattice simulation for supersymmetric gauge theories in one
  dimension,''
Phys.\ Rev.\ Lett.\  {\bf 99}, 161602 (2007), 
  arXiv:0706.1647 [hep-lat].




\bibitem{AHNT}
  K.~N.~Anagnostopoulos,
  M.~Hanada, J.~Nishimura and S.~Takeuchi,
\emph{Monte Carlo studies of 
supersymmetric matrix quantum mechanics 
with sixteen supercharges at finite temperature},
Phys.\ Rev.\ Lett.\  {\bf 100} (2008) 021601,  
[{\tt arXiv:0707.4454[hep-th]}].

  M.~Hanada, A.~Miwa, J.~Nishimura and S.~Takeuchi,
  ``Schwarzschild radius from Monte Carlo calculation of the Wilson loop in
  supersymmetric matrix quantum mechanics,''
  Phys.\ Rev.\ Lett.\  {\bf 102}, 181602 (2009), 
  [arXiv:0811.2081 [hep-th]].
  
  M.~Hanada, Y.~Hyakutake, J.~Nishimura and S.~Takeuchi,
   ``Higher derivative corrections to black hole thermodynamics from
  supersymmetric matrix quantum mechanics,''
  Phys.\ Rev.\ Lett.\  {\bf 102}, 191602 (2009),   
  [arXiv:0811.3102 [hep-th]].
  
  M.~Hanada, J.~Nishimura, Y.~Sekino and T.~Yoneya,
  ``Monte Carlo studies of Matrix theory correlation functions,''
  Phys.\ Rev.\ Lett.\ {\bf 104}, 151601 (2010). 


\bibitem{Catterall-Wiseman}
  S.~Catterall and T.~Wiseman,
``Black hole thermodynamics from simulations 
 of lattice Yang-Mills theory,''
  Phys.\ Rev.\  D {\bf 78}, 041502 (2008)
  [arXiv:0803.4273 [hep-th]].
  
S.~Catterall and T.~Wiseman,
  ``Extracting black hole physics from the lattice,''
  arXiv:0909.4947 [hep-th].

\bibitem{BMN}
  D.~E.~Berenstein, J.~M.~Maldacena and H.~S.~Nastase,
  ``Strings in flat space and pp waves from N = 4 super Yang Mills,''
  JHEP {\bf 0204}, 013 (2002)  
  [arXiv:hep-th/0202021].

\bibitem{Myers:1999ps}
  R.~C.~Myers,
  ``Dielectric-branes,''
  JHEP {\bf 9912}, 022 (1999). 
  [arXiv:hep-th/9910053].


\bibitem{Maldacena:2002rb}
  J.~M.~Maldacena, M.~M.~Sheikh-Jabbari and M.~Van Raamsdonk,
  ``Transverse fivebranes in matrix theory,''
  JHEP {\bf 0301}, 038 (2003). 


\bibitem{Ishii:2008ib}
  T.~Ishii, G.~Ishiki, S.~Shimasaki and A.~Tsuchiya,
  ``N=4 Super Yang-Mills from the Plane Wave Matrix Model,''
  Phys.\ Rev.\  D {\bf 78}, 106001 (2008). 
  [arXiv:0807.2352 [hep-th]].
  
  
  G.~Ishiki, S.~W.~Kim, J.~Nishimura and A.~Tsuchiya,
  ``Deconfinement phase transition in N=4 super Yang-Mills theory on $R\times S^3$ from
  supersymmetric matrix quantum mechanics,''
  Phys.\ Rev.\ Lett.\  {\bf 102}, 111601 (2009)
  [arXiv:0810.2884 [hep-th]].

  
  
  G.~Ishiki, S.~W.~Kim, J.~Nishimura and A.~Tsuchiya,
  ``Testing a novel large-N reduction for N=4 super Yang-Mills theory on
  $R\times S^3$,''
  JHEP {\bf 0909}, 029 (2009)
  [arXiv:0907.1488 [hep-th]].

  

\bibitem{Hanada:2009hd}
  M.~Hanada, L.~Mannelli and Y.~Matsuo,
  ``Four-dimensional N=1 super Yang-Mills from matrix model,''
  Phys.\ Rev.\  D {\bf 80}, 125001 (2009)
  [arXiv:0905.2995 [hep-th]].

  M.~Hanada, L.~Mannelli and Y.~Matsuo,
  ``Large-N reduced models of supersymmetric quiver, Chern-Simons gauge
  theories and ABJM,''
  JHEP {\bf 0911}, 087 (2009)  
  [arXiv:0907.4937 [hep-th]].
  

\bibitem{Eguchi:1982nm}
  T.~Eguchi and H.~Kawai,
  ``Reduction Of Dynamical Degrees Of Freedom In The Large N Gauge Theory,''
  Phys.\ Rev.\ Lett.\  {\bf 48}, 1063 (1982).


\bibitem{Hanada:2010kt}
  M.~Hanada, S.~Matsuura, F.~Sugino,
  ``Two-dimensional lattice for four-dimensional N=4 supersymmetric Yang-Mills,''
  Prog.\  Theor.\  Phys.\  {\bf 126}, 597-611 (2011).
  [arXiv:1004.5513 [hep-lat]].



\bibitem{Sugino:2003yb}
  F.~Sugino,
  ``A lattice formulation of super Yang-Mills theories with exact
  supersymmetry,''
  JHEP {\bf 0401}, 015 (2004)  
  [arXiv:hep-lat/0311021].
  
\bibitem{Sugino:2004qd}
  F.~Sugino,
  ``Super Yang-Mills theories on the two-dimensional lattice with exact
  supersymmetry,''
  JHEP {\bf 0403}, 067 (2004)  
  [arXiv:hep-lat/0401017].

\bibitem{Sugino:2004uv}
  F.~Sugino,
  ``Various super Yang-Mills theories with exact supersymmetry on the
  lattice,''
  JHEP {\bf 0501}, 016 (2005)  
  [arXiv:hep-lat/0410035].

\bibitem{Catterall_private}
  S.~Catterall, private communication. 

\bibitem{Kaplan:2002wv}
  D.~B.~Kaplan, E.~Katz and M.~Unsal,
  ``Supersymmetry on a spatial lattice,''
  JHEP {\bf 0305}, 037 (2003)  
  [arXiv:hep-lat/0206019].

\bibitem{Cohen:2003xe}
  A.~G.~Cohen, D.~B.~Kaplan, E.~Katz and M.~Unsal,
  ``Supersymmetry on a Euclidean spacetime lattice. I: A target theory with
  four supercharges,''
  JHEP {\bf 0308}, 024 (2003)  
  [arXiv:hep-lat/0302017].
  
	  A.~G.~Cohen, D.~B.~Kaplan, E.~Katz and M.~Unsal,
  ``Supersymmetry on a Euclidean spacetime lattice. II: Target theories  with
  eight supercharges,''
  JHEP 
ibid. {\bf 0312}, 031 (2003)  
  [arXiv:hep-lat/0307012].


\bibitem{KaplanUnsal16SUSY}  
  D.~B.~Kaplan and M.~Unsal,
  ``A Euclidean lattice construction of supersymmetric Yang-Mills theories
  with sixteen supercharges,''
  JHEP {\bf 0509}, 042 (2005) 
  [arXiv:hep-lat/0503039].


\bibitem{Catterall:2004np}
  S.~Catterall,
  ``A geometrical approach to N = 2 super Yang-Mills theory on the two
  dimensional lattice,''
  JHEP {\bf 0411}, 006 (2004)  
  [arXiv:hep-lat/0410052].


  S.~Catterall,
  ``Lattice formulation of N = 4 super Yang-Mills theory,''
  JHEP {\bf 0506}, 027 (2005)
  [arXiv:hep-lat/0503036].

\bibitem{D'Adda:2005zk}
  A.~D'Adda, I.~Kanamori, N.~Kawamoto and K.~Nagata,
  ``Exact extended supersymmetry on a lattice: Twisted N = 2 super  Yang-Mills
  in two dimensions,''
  Phys.\ Lett.\  B {\bf 633}, 645 (2006)  
  [arXiv:hep-lat/0507029].


\bibitem{Damgaard:2007be}
  P.~H.~Damgaard and S.~Matsuura,
  ``Classification of Supersymmetric Lattice Gauge Theories by Orbifolding,''
  JHEP {\bf 0707}, 051 (2007)
  [arXiv:0704.2696 [hep-lat]].


\bibitem{Unsal:2006qp}
  M.~Unsal,
  ``Twisted supersymmetric gauge theories and orbifold lattices,''
  JHEP {\bf 0610}, 089 (2006)
  [arXiv:hep-th/0603046].

  T.~Takimi,
  ``Relationship between various supersymmetric lattice models,''
  JHEP {\bf 0707}, 010 (2007)
  [arXiv:0705.3831 [hep-lat]].

  P.~H.~Damgaard and S.~Matsuura,
  ``Relations among Supersymmetric Lattice Gauge Theories via Orbifolding,''
  JHEP {\bf 0708}, 087 (2007)
  [arXiv:0706.3007 [hep-lat]].


\bibitem{Catterall:2007kn}
  S.~Catterall,
  ``From Twisted Supersymmetry to Orbifold Lattices,''
  JHEP {\bf 0801}, 048 (2008)
  [arXiv:0712.2532 [hep-th]].

 

\bibitem{Hashimoto:1999ut} 
  A.~Hashimoto and N.~Itzhaki,
  ``Non-commutative Yang-Mills and the AdS/CFT correspondence,''
  Phys.\ Lett.\  B {\bf 465}, 142 (1999)  
  [arXiv:hep-th/9907166].
  
  J.~M.~Maldacena and J.~G.~Russo,
  ``Large N limit of non-commutative gauge theories,''
  JHEP {\bf 9909}, 025 (1999) 
  [arXiv:hep-th/9908134]. 

\bibitem{Matusis:2000jf}
  A.~Matusis, L.~Susskind and N.~Toumbas,
  ``The IR/UV connection in the non-commutative gauge theories,''
  JHEP {\bf 0012}, 002 (2000)  
  [arXiv:hep-th/0002075].


\bibitem{Das:2003yq}
  K.~Sugiyama and K.~Yoshida,
  ``Type IIA string and matrix string on pp-wave,''
  Nucl.\ Phys.\  B {\bf 644}, 128 (2002)  
  [arXiv:hep-th/0208029].

  S.~R.~Das, J.~Michelson and A.~D.~Shapere,
  ``Fuzzy spheres in pp-wave matrix string theory,''
  Phys.\ Rev.\  D {\bf 70}, 026004 (2004) 
  [arXiv:hep-th/0306270].
 
  G.~Bonelli,
  ``Matrix strings in pp-wave backgrounds from deformed super Yang-Mills
  theory,''
  JHEP {\bf 0208}, 022 (2002)   
  [arXiv:hep-th/0205213].
  
\bibitem{Unsal:2005us}
  M.~Unsal,
  ``Supersymmetric deformations of type IIB matrix model as matrix
  regularization of N = 4 SYM,''
  JHEP {\bf 0604}, 002 (2006)
  [arXiv:hep-th/0510004].


\bibitem{Ydri:2007ua}
  B.~Ydri,
  ``A Proposal for a Non-Perturbative Regularization of {\cal N}=2 SUSY 4D
  Gauge Theory,''
  Mod.\ Phys.\ Lett.\  A {\bf 22}, 2565 (2007)
  [arXiv:0708.3066 [hep-th]].

\bibitem{Bal:2004ai}
  S.~Bal, M.~Hanada, H.~Kawai and F.~Kubo,
  ``Fuzzy torus in matrix model,''
  Nucl.\ Phys.\  B {\bf 727}, 196 (2005)
  [arXiv:hep-th/0412303].

  T.~Azeyanagi, M.~Hanada and T.~Hirata,
  ``On Matrix Model Formulations of Noncommutative Yang-Mills Theories,''
  Phys.\ Rev.\  D {\bf 78}, 105017 (2008)
  [arXiv:0806.3252 [hep-th]].

\bibitem{Kaneko:2005kp}
  H.~Kaneko, Y.~Kitazawa and D.~Tomino,
  ``Fuzzy spacetime with SU(3) isometry in IIB matrix model,''
  Phys.\ Rev.\  D {\bf 73}, 066001 (2006)
  [arXiv:hep-th/0510263].
  
\bibitem{ArkaniHamed:2001ca}
  N.~Arkani-Hamed, A.~G.~Cohen and H.~Georgi,
  ``(De)constructing dimensions,''
  Phys.\ Rev.\ Lett.\  {\bf 86}, 4757 (2001)
  [arXiv:hep-th/0104005].
  
 
\bibitem{Hanada:2009hq}
  M.~Hanada and I.~Kanamori,
  ``Lattice study of two-dimensional ${\cal N}=(2,2)$ super Yang-Mills at large-$N$,''
  Phys.\ Rev.\  D {\bf 80}, 065014 (2009). 
  [arXiv:0907.4966 [hep-lat]].
  
  
\bibitem{Aoki:1999vr}
  H.~Aoki, N.~Ishibashi, S.~Iso, H.~Kawai, Y.~Kitazawa and T.~Tada,
  ``Noncommutative Yang-Mills in IIB matrix model,''
  Nucl.\ Phys.\  B {\bf 565}, 176 (2000)
  [arXiv:hep-th/9908141].

  J.~Madore, S.~Schraml, P.~Schupp and J.~Wess,
  ``Gauge theory on noncommutative spaces,''
  Eur.\ Phys.\ J.\  C {\bf 16}, 161 (2000)
  [arXiv:hep-th/0001203].


\bibitem{Gregory:1993vy}
  R.~Gregory and R.~Laflamme,
  ``Black strings and p-branes are unstable,''
  Phys.\ Rev.\ Lett.\  {\bf 70}, 2837 (1993)
  [arXiv:hep-th/9301052].

\bibitem{Aharony:2004ig}
See e.g. 
  O.~Aharony, J.~Marsano, S.~Minwalla and T.~Wiseman,
  ``Black hole-black string phase transitions in thermal 1+1-dimensional
  supersymmetric Yang-Mills theory on a circle,''
  Class.\ Quant.\ Grav.\  {\bf 21}, 5169 (2004)
  [arXiv:hep-th/0406210].




\bibitem{Kawahara:2007fn}
  N.~Kawahara, J.~Nishimura and S.~Takeuchi,
  ``Phase structure of matrix quantum mechanics at finite temperature,''
  JHEP {\bf 0710}, 097 (2007)
  [arXiv:0706.3517 [hep-th]].

\bibitem{Azeyanagi:2009zf}
  T.~Azeyanagi, M.~Hanada, T.~Hirata and H.~Shimada,
  ``On the shape of a D-brane bound state and its topology change,''
  JHEP {\bf 0903}, 121 (2009)
  [arXiv:0901.4073 [hep-th]].

\bibitem{Catterall:2010fx}
  S.~Catterall, A.~Joseph and T.~Wiseman,
  ``Thermal phases of D1-branes on a circle from lattice super Yang-Mills,''
  arXiv:1008.4964 [hep-th].

\end{thebibliography}
\end{document}